# Embedding Spatial Software Visualization in the IDE: an Exploratory Study

Adrian Kuhn
Software Composition Group
University of Bern
akuhn@iam.unibe.com

David Erni
Software Composition Group
University of Bern
deif@students.unibe.com

Oscar Nierstrasz
Software Composition Group
University of Bern
oscar@iam.unibe.com

## ABSTRACT

*Software visualization can be of great use for understanding and exploring a software system in an intuitive manner. Spatial representation of software is a promising approach of increasing interest. However, little is known about how developers interact with spatial visualizations that are embedded in the IDE. In this paper, we present a pilot study that explores the use of Software Cartography for program comprehension of an unknown system. We investigated whether developers establish a spatial memory of the system, whether clustering by topic offers a sound base layout, and how developers interact with maps. We report our results in the form of observations, hypotheses, and implications. Key findings are a) that developers made good use of the map to inspect search results and call graphs, and b) that developers found the base layout surprising and often confusing. We conclude with concrete advice for the design of embedded software maps.*

## 1. INTRODUCTION

Software visualization can be of great use for understanding and exploring a software system in an intuitive manner. In the past decade the software visualization community has developed a rich wealth of visualization approaches [12] and provided evidence of their usefulness for expert tasks, such as reverse engineering, release management or dynamic analysis (*e.g.*, [37, 8, 32, 29]). Typically, these visualization approaches had been implemented in interactive tools [33]. However most of these tools are stand-alone prototypes that have never been integrated in an IDE (integrated development environment). Little is thus known about the benefits of software visualization for the "end users" in software engineering, that is for everyday programmers. What is lacking is how these techniques support the day to day activities of software developers [36].

In this paper, we report on a pilot study of a spatial software visualization that is embedded in the IDE. The spatial visualization is based on the Software Cartography approach that has been presented and introduced in previous work [22, 20, 15]. Spatial representation of software is a promising research field of increasing interest [39, 11, 5, 35, 24, 27], however the respective tools are either not tightly integrated in an IDE or have not yet been evaluated in a user study. Spatial representation of software is supposed to support developers in establishing a long term, spatial memory of the software system. Developers may use spatial memory to recall the location of software artifacts, and to put thematic map overlays in relation with each other [20].

The scenario of our user study is first contact with an unknown closed-source system. Our main question was whether and how developers make use of the embedded visualization and if our initial assumptions made when designing the visualization (as for example the choice of lexical similarity as the map's base layout [20, Sec 3]) are based on a valid model of developer needs. Participants had 90 minutes to solve 5 exploratory tasks and to fix one bug report. We used the think-aloud protocol and recorded the voices of the participants together with a screen capture of their IDE interactions. We took manual notes of IDE interaction sequences and annotated the sequences with the recorded think-aloud transcripts.

Results are mixed — some support and some challenge our assumptions on how developers would use the embedded visualization. Participants found the map most useful to explore search results and call graphs, but only rarely used the map for direct navigation as we would have expected.

Contributions of this paper are as follows:

- We embedded the stand-alone CODEMAP prototype in the Eclipse IDE, and added novel thematic overlays that support the most important development tasks with visual feedback (see Section 2 and Section 3).

- We performed a think-aloud user study to evaluate the use of spatial visualization in the IDE. We discuss and comment on our results, and conclude with practical design implications (see Section 6 and Section 7).

- We provide suggestions on how to improve Software Cartography and the CODEMAP tool (see Section 8).

The remainder of this paper is structured as follows: Section 2 and Section 3 present the new CODEMAP prototype; Section 4 and Section 5 describes design and data analysis of the study; Section 6 is the main part that presents and discusses the results; Section 7 explains threats to the validity; Section 8 discusses related work; Section 9 concludes.

## 2. SOFTWARE CARTOGRAPHY

Software Cartography uses a spatial visualization of software systems to provide software development teams with a stable and shared mental model. The basic idea of cartographic visualization is to apply thematic cartography [34] on software visualization. That is, to show thematic overlays on top of a stable, spatial base layout. Features on a thematic map are either point-based, arrow-based or continuous. For software this could be the dispersion of design flaws as visualized using icons; a call graph is visualized as a flow map (as illustrated on Figure 1); and test coverage is visualized as a choropleth map, *i.e.*, a heat map.

Software Cartography is most useful when it supports as many development tasks with spatial location awareness as possible. We therefore integrated our prototype into the Eclipse IDE so that a map of the software system may always be present. This helps developers to correlate as many development tasks as possible with their spatial location.

At the moment, the CODEMAP plug-in for Eclipse supports the following tasks:[1]

- Navigation within a software system, be it for development or analysis. CODEMAP is integrated with the package explorer and editor of Eclipse. The selection in the package explorer and the selection on the map are linked. Open files are marked with an icon on the map. Double clicking on the map opens the closest file in the editor. When using heat map mode, recently visited classes are highlighted on the map.

- Comparing software metrics to each other, *e.g.*, to compare bug density with code coverage. The map displays search results, compiler errors, and (given the Eclemma plug-in is installed) test coverage information. More information can be added through an plug-in extension point.

- Social awareness of collaboration in the development team. CODEMAP can connect two or more Eclipse instances to show open files of other developers. Colored icons are used to show the currently open files of all developers. Icons are colored by user and updated in real time.

- Understand a software system's domain. The layout of CODEMAP is based on clustering software by topic [19], as it has been shown that, over time, the lexicon of source code is more stable than its structure [1]. Labels on the map are not limited to class names, but include automatically retrieved keywords and topics.

- Exploring a system during reverse engineering. CODEMAP is integrated with Eclipse's structural navigation features, such as search for callers, implementers, and references. Arrows are shown for search results. We apply the FLOW MAP algorithm [30] to avoid visual clutter by merging parallel arrow edges. Figure 1 shows the result of searching for calls to the `#getSettingOrDefault` method in the `MenuAction` class .

---
[1]http://scg.unibe.ch/codemap

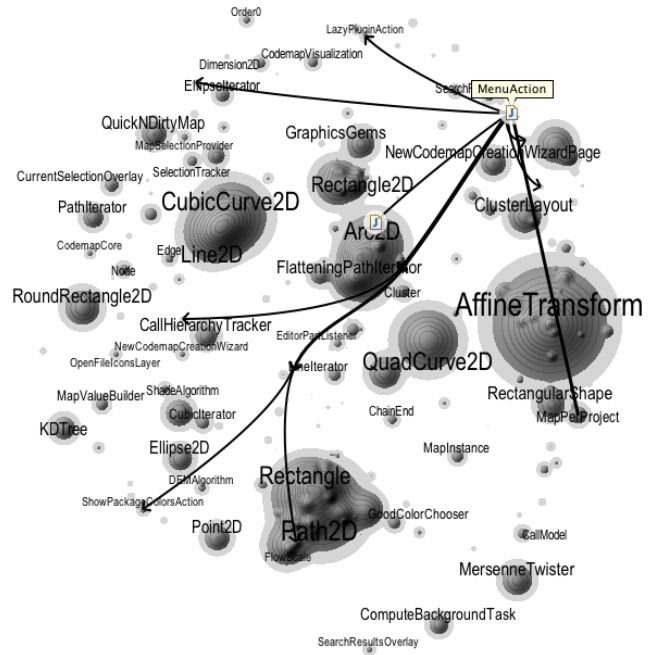

**Figure 1:** *Thematic codemap of a software system. Here the* Codemap *tool itself is shown. Arrow edges show incoming calls to the* `#getSettingOrDefault` *method in the* `MenuAction` *class, which is currently active in the editor and thus labeled with a pop-up.*

## 3. THE CODEMAP ALGORITHM

Figure 2 illustrates the construction of a software map. The sequence of the construction is basically the same as presented in previous work [21, 20].

*2-Dimensional Embedding.*

A distance metric is used to compute the pair-wise dissimilarity of software artifacts (typically source code files). A combination of the Isomap algorithm [38] and Multidimensional Scaling (MDS) [4] is used to embed all software artifacts into the visualization pane. The application of Isomap is an improvement over previous work in order to assist MDS with the global layout. In contrast to our previous work, Latent Semantic Indexing (LSI) is not applied anymore, it has been found to have little impact on the final embedding.

*Digital Elevation Model.*

In the next step, a digital elevation model is created. Each software artifact contributes a Gaussian shaped basis function to the elevation model according to its KLOC size. The contributions of all software artifacts are summed up and normalized.

*Cartographic rendering.*

In the final step, hill-shading is used to render the landscape of the software map. Please refer to previous work for full details [21, 20]. Metrics and markers are rendered in transparent layers on top of the landscape. Users can toggle separate layers on/off and thus customize the codemap display to their needs.

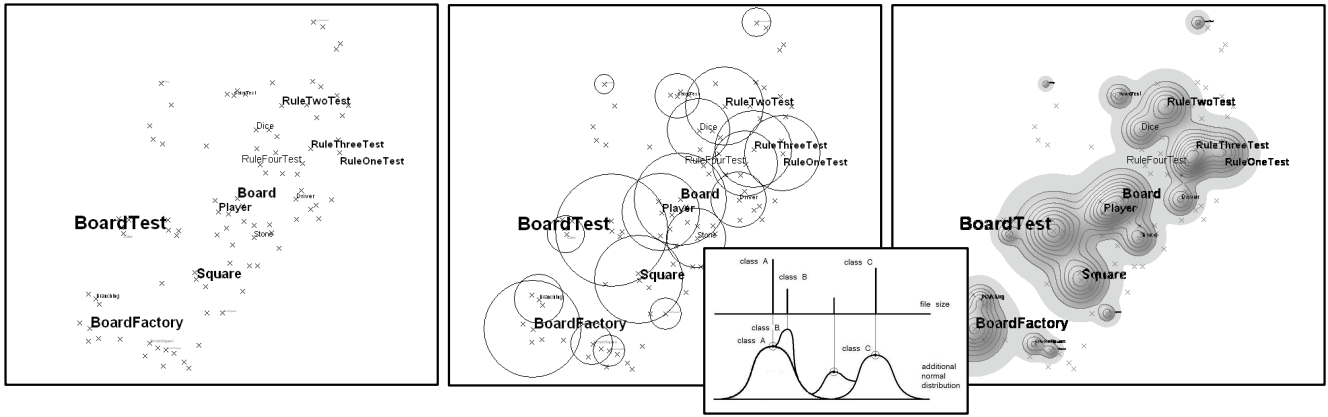

**Figure 2:** *Construction steps of a software map, from left to right: 1) 2-dimensional embedding of files on the visualization pane; 2.a) circles around each file's location, based on class size in KLOC; 2.b) each file contributes a Gaussian shaped basis function to the elevation model according to its KLOC size; the contributions of all files are summed up; 3) fully rendered map with hill-shading, contour lines, and filename labels.*

## 4. METHODOLOGY

We evaluated our approach in a pilot study with professional developers and students. The scenario investigated by the experiment is first contact with an unknown software system. Participants have 90 minutes to solve 5 program comprehension tasks and to fix one bug report. After the experiment, participants are asked to sketch a drawing of their mental map of the system.

Our goal for the present pilot study was to learn about the usability of CODEMAP for program comprehension. We have been seeking to answer several questions. How can we support developers in establishing a spatial memory of software systems? How do we best support the developers spatial memory using software visualization? How to best embed spatial software visualization in the IDE? When provided with spatial representation of search results and call graphs, how do developers make use of them?

Not covered in this study, and thus open for future user studies, are the shared team awareness and long term memory claims of the Software Cartography approach [20].

### 4.1 Design of the Study

The study consists of six programming tasks. The training task introduced the participants to the CODEMAP plug-in. The first five tasks were program comprehension tasks, starting with general questions and then going into more and more detailed questions. Eventually, the last task was to fix an actual bug in the system. Participants were asked to use the map whenever they saw fit, but otherwise they were free to use any other feature of Eclipse they wanted.

*Task 1, Domain and Collaborators.*
"Find the purpose of the given application and identify the main collaborators. Explore the system, determine its domain, and fulfil the following tasks: a) describe the domain, b) list the main collaborators, c) draw a simple collaboration diagram, d) identify the main feature of the application."

*Task 2, Technologies.*
"In this task we are interested in the technologies used in the application. List the main technologies, such as for example Ajax, XML, or unit testing."

*Task 3, Architecture.*
"In this task we are going to take a look at the architecture of the application. Reverse engineer the architecture by answering the following questions: a) which architectural paradigm is used (as for example pipes and filters, layers, big ball of mud, etc)? b) what are the main architectural components? c) how are those components related to one another? d) draw a UML diagram at the level of components."

*Task 4, Feature Location.*
"In this task we are interested in classes that collaborate in a given feature. Please locate the following features: a) Interactive users are reminded after some months, and eventually deleted if they do not log in after a certain number of months, b) Depending on the kind of user, a user can see and edit more or less data. There are permission settings for each kind of user that are checked whenever data is accesses, and c) Active search: the system compares the curriculum vitae of the users with stored searches of the companies and mails new matches to the companies."

*Task 5, Code Assessment.*
"In this task we want to assess the code quality of the application. Please answer the following questions: a) what is the degree of test coverage? b) are there any god classes? c) are the classes organized in their proper packages? Should certain classes be moved to other packages? Please list two to three examples."

We provided a code coverage plug-in with the experiment, as well as a definition of what constitutes a god class [23].

*Task 6, Bug Fixing.*
In this task we provided an actual bug report and asked "Describe how you would handle the bug report, that is how and where you would change the system and which classes are involved in the bug fix. You are not asked to actually fix the bug, but just to describe how you would fix it."

## 4.2 Participant Selection

Participants were selected through an open call for participation on Twitter[2] as well as through flyers distributed at a local Eclipse event. Subjects were required to be medium level Java programmers with at least one year of experience with both Java and Eclipse programming. The six tasks had been designed so that the participants did not need to be knowledgeable with the provided application, but rather that they explore it as they go along. Seven participants took part in the experiment: 4 graduate students and 3 professional developers from industry. None of the participants was familiar with the provided application or with the Codemap plugin; even though some had attended a 15 minute presentation about the Codemap plugin at the Eclipse event mentioned above.

## 4.3 Study Setting

The study consisted of three main parts. The first part was the training task in which the participants were given a short presentation of Codemap and a tutorial document that explained all features of the Codemap plug-in. The tutorial explained all features mentioned in Section 2 using walk-through descriptions of their use. The participants were given 20 minutes to explore a small example program using the Codemap plug-in. When they felt ready, we started part two of the experiment.

The second part consisted of the actual programming tasks. A fixed amount of time was allotted to each task. Participants were asked to spend no more than 15 minutes on each task. All subjects had access to the Codemap plugin as our aim was to explore their use of the plugin rather than to compare a controlled parameter against the baseline.

Eventually, in a third part we held a debriefing session. We asked participants to draw a map (with any layout or diagram language whatsoever) of how they would explain the system under study to another developer. We asked the participants for feedback regarding their use of the Codemap plugin and how the plugin could be improved.

## 5. DATA COLLECTION

We asked the participants to think aloud, and recorded their voice together with a captured video of their computer screen using the Camtasia software[3]. We reminded the participants to think aloud whenever they fell silent: we told them to imagine a junior programmer sitting beside them to whom they are to explain their actions (Master/Apprentice [2]). The participants were asked to respond to a survey while performing the study. The survey consisted of their answers to the tasks, as well as the perceived difficulty of the tasks and whether they found the Codemap plugin useful for the task at hand. We used a combination of semantic differential statements and Likert scales with a 5 point scale.

We measured whether or not subjects were successful in completing a programming task. We used three success levels to measure the success and failure of tasks: a task could be a success, a partial success or a failure. We further subdivided tasks 4 and 5 into three subtasks and recorded success levels for each individual subtask. We asked one of the original authors of the system to assess the success levels. As this was a think-aloud study, we did not measure time, but alloted a fixed 15 minute slot to each task.

Our main interest was focused on how the participants used the IDE to solve the tasks, independent of their success level. To do this, we transcribed important quotes from the recorded participant voices and screen captures and took notes of the actions that the participants did during the tasks. For each task we tracked the use of the following IDE elements:

- Browsing the system using the *Package Explorer* and *Outline* view. This includes both drill-down as well as linear browsing of package, class and method names.

- Browsing the system using the spatial visualization of the Codemap plugin. This includes both opening single classes, selecting a whole cluster of classes on the map, as well as reading class name labels on the map.

- Reading source code in the editor pane, including documentation in the comments of class and method headers.

- Navigating the structure of the system using the *Type Hierarchy* and *Call Hierarchy* view. We tracked whether they explored the results of these searches in Eclipse's tabular result view or using the flow-map arrows displayed on the spatial visualization of Codemap.

- Searching the structure of the system with either the *Open Type* or *Java Search* dialog. This allows users to search for specific structural elements such as classes, methods or fields. Again, we tracked whether they explored the results in Eclipse's result view or on the visualization of Codemap.

- Searching the system with the unstructured text search, either through the *Java Search* dialog or the immediate search bar of the Codemap plugin. Also here, we tracked whether they explored the results in Eclipse's result view or on the visualization of Codemap.

Replicability: the raw data of our analysis is available on the Codemap website at `http://scg.unibe.ch/codemap`.

## 6. RESULTS

After analyzing our data, we observed different degrees of interaction with the Codemap plug-in. We focused our analysis on interaction sequences that included interaction with the Codemap plug-in, but also on those interaction sequences that challenged our assumptions about how developers would make use of the plug-in.

The presentation of results is structured as follows. First, we briefly cover how each task was solved. Then present an in-depth analysis of our observations, structured by triples of *observation*, *hypothesis*, and *implication*. Implications are directed at improving the design and usability of spatial visualizations that are embedded in an IDE.

## 6.1 Task Performance

*Task 1, Domain and Collaborators.*
Participants used an approach best described as a "reverse Booch method" [3]. Given a two-sentence description of the system that we've provided, they searched for nouns and verbs using Eclipse's full text search. Most participants used CODEMAP to assess quantity and dispersion of search results, and also to directly select and inspect large classes. Then they looked at the class names of the matches to learn about the domain and collaborators of the system. Students also read source code, whereas professional participants limited their investigation to using the package explorer and class outline.

---
[2] `http://twitter.com/codemap`
[3] `http://www.techsmith.com/camtasia`

*Task 2, Technologies.*

This task showed the most uniform behavior from both student and professional participants. They inspected the build path node and opened all included JAR libraries. Professional developers typically raised the concern that possibly not all of these libraries were (still) used and started to explore whether they were used. Typically they would carry out a search to do so, but one developer showed a very interesting pattern: He would remove the libary "on purpose" and then look for compile errors as an indicator of its use. Students seems to implicitly assume that all libraries were actually used, at least they never raised such a concern. We interpret this as a sign that professionals are more cautious [18] and thus more aware of the typical decay caused by software evolution, which may include dead libraries.

*Task 3, Architecture.*

Typically participants drilled-down with the package explorer and read all package names. All professionals started out by formulating the hypothesis of a layered three-tier architecture, and then start fitting the packages to the different layers. Most participants used CODEMAP to look at the dispersion of a package's classes (when selecting a package in the package explorer, the contained classes are highlighted on the map).

To learn about the architectural constraints, professionals, for the first time in the experiment, started reading source code. They also did so quite differently from the way that students did. Whereas students typically read code line by line, trying to understand what it does, the professionals rather used the scroll-wheel to skim over the code as it flies by on the screen, thereby looking for "landmarks" such as constructor calls, method signatures and field definitions. Professionals made much more use of "open call hierarchy" and "open type hierarchy". Interestingly enough, only one participant opened a type hierarchy of the whole project.

*Task 4, Feature Location.*

For this task, participants made most frequent and more interesting use of CODEMAP than for any other task. Same of for task 1, participants used a reversal of the Booch method. They searched for nouns and verbs found in the feature description. Again, they used the map to assess quantity and dispersion of search results. Also two participants used the map to select and inspect search matches based on their context in the map.

Participants now began to read more source code than before. In particular, when they found a promising search result they used the "open call hierarchy" feature to locate related classes. All participants reported that CODEMAP flowmap overlay helped them to work with the call graph. For some developers there was an actual "Aha moment" where one glance at the CODEMAP helped them to solve the current subtask immediately without further investigation. Figure 3 illustrates one particular moment as encountered by participant T during the location of the security feature.

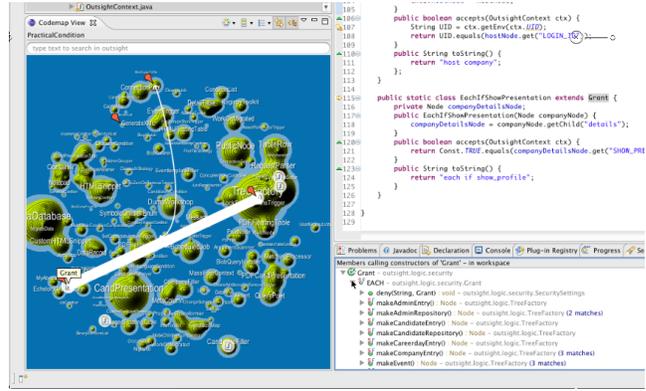

**Figure 3:** *Screen capture of "Aha moment" as encountered by participant T during task 4-b (location of security features): Upon opening the call hierarchy of* Grant*'s constructor, a huge call-arrow appeared on the map: indicating dozens of individual calls that connect the security-related archipelago in the southwest with the* TreeFactory *island in the east. Given the visual evidence of this arrow, participant T solved the task without further investigation.*

*Task 5, Code Assesment.*

This set of tasks made it most obvious that CODEMAP's layout was not based on package structure. Participants reported that they had a hard time to interpret the thematic maps as they could not map locations on the map to packages. In particular the professional participants expressed concerns regarding the use of KLOC for hill size. They expressed concerns that this might be misleading since lines of code is not always an indicator of importance or centrality on the system's design.

*Task 6, Bug Fixing.*

Participants mainly used the same approach as for the feature location tasks. They first located the implementation of the feature in which the bug occurs, and then fixed the bug. Professional participants did so successfully, whereas student participants did not manage to find the correct location in the source code.

*Wrap-up session.*

In general, participants reported that CODEMAP was most useful when it displayed search results, callers, implementers, and references. A participant reported: *"I found it very helpful that you get a visual clue of quantity and distribution of your search results"*. In fact, we observed that that participants rarely used the map for direct navigation but often for search and reverse engineering tasks.

Another observation was that inexperienced developers (*i.e.*, students) are more likely to find the map useful than professional developers. This might be explained by the hypothesis that to power users *any* new way of using the IDE is likely to slow them down, and conversely to beginners *any* way of using the IDE is novel. The only exception to this ob-servation was CODEMAP's search bar, a one-click interface to Eclipse's native search, that was appreciated and used by all participants but one that preferred to use the search dialog.

One Participant also provided us feedback comparing his experience with CODEMAP to that with the Moose analysis tool [26]. He uses Moose at work after having attended a tutorial by a consultant. He said he prefers the immediate feedback of Codemap, and reported that *"the gap between Moose and IDE is just too large, not to mention the struggle of importing Java code. Moose helps you to e.g., find god-classes but this is typically not new to developers that know a system. Codemap seems more interesting as it integrates with what you actually do in the IDE as you program."*

## 6.2 Observations, Hypotheses, Implications

In this section, we present an in-depth analysis of our observations, structured by triples of *observation*, *hypothesis*, and *implication*. Implications are directed at improving the design and usability of spatial visualizations that are embedded in an IDE.

*Observation 6.1: When thinking aloud, developers did speak of the system's architecture in spatial terms.*

The think-aloud protocol revealed that participants refer to the system's architecture in spatial terms. Professional participants referred to packages as being above, below, or at the some level as one another. Some of them even did so before recovering the system's 3-tier architecture in task #3. Most professionals referred to utility packages a being spatially beside or outside the layered architecture.

For example, participant T located all utility packages in the upper left corner, separated by a jagged line. While doing so, he made a gesture as if pushing the utility classes away and stated, *"I am putting them up here because to me they are somehow beside the system."*

Students on the other hand made much fewer references to the system's architecture, both spatial as well as in general. They were typically reasoning about the system at the level of classes and source lines, rather than in architectural terms. The maps drawn by students in the wrap-up phase, however, showed similar spatial structure to those of the professionals. It remains thus open whether students established a genuine spatial model while working with the code (as we observed for professionals) or only because they were asked to draw the wrap-up maps.

*Hypothesis 6.1: Professional developers do establish a spatial mental model of the system's architecture.*

Based on above observations there is evidence to assume that professional developers establish a spatial mental model of the system's architecture as they work with code. Furthermore, they do so even without visual aids, since they use spatial terms and thinking even before being asked to draw a diagram of the system's architecture.

*Implication 6.1: Developers should be able to arrange the layout according to their mental model.*

This has implications on the design of a system's spatial visualization. Developers should be able to arrange the layout according to their mental model. Developers should be able to drag and move parts of the map around as they wish, rather than having to stick with the automatically established layout. Code Canvas [11] and Code Bubbles [5] both already address this implication. In those tools, the user may drag individual elements around and arrange them according to his mental model.

We observed that developers referred to architectural components, but not classes, in spatial terms. The needs of developers might thus be even better served by providing them more high-level means of arranging the map. Our next prototype will use *anchored multidimensional scaling* such that developers may initialize the map to their mental model. Anchored MDS allows the developer to define anchors which influence the layout of the map [6, Sec 4.4]. Any software artifact can be used as an anchor (as long as we can compute a its distance to artifacts on the map), even for example external libraries. In this way, developers might *e.g.*, arrange the database layer in the south and the UI layer in the north using the respective libraries as anchors.

*Observation 6.2: Participants used Codemap to assess quantity and dispersion of search results and call graphs.*

The feature of Codemap that was used most often, by both professionals and students, was the illustration of search results and call graphs. Participants reported that they liked the search-result support of the map, explaining that it gives them much faster initial feedback than Eclipse's tabular presentation of search results. Many participants reported that it was *"as if you could feel the search results,"* and that *"you get an immediate estimate how much was found, whether it is all one place or scattered all over the place."*

Figure 3 illustrates one particular "Aha moment" as encountered by participant T during task 4-b, *i.e.*, location of security features: Upon opening the call hierarchy, a huge call-arrow appeared on the map: indicating dozens of individual calls that connect the security-related archipelago in the south-west with the `TreeFactory` island in the east. Given the visual evidence of this arrow, the participant solved the task immediately without further investigation of the system.

*Hypothesis 6.2: Intuitive visualization to show quantity and dispersion of search results (as well as call graphs) address an important need of developers.*

Given the above observation it seems clear that developers have urgent needs for better representation of search results than tabular lists. We found that both students and professionals used the map to get an immediate estimation of search results. This is most interesting since otherwise their use of the tabular search results differed: Professionals glanced at the results, inspected one or maybe two results, and then either accepted or rejected their hypothesis about the system, while students would resort to a linear search through all search results, not daring to reject a hypothesis on the grounds of one or two inspected results only.

Given the map's illustration of search results however, the behavior of both groups changed. Students dared to take quick decisions from a mere glance at the map, whereas professionals were more likely to inspect several results. One professional reported that he *"inspected more results than usual, because the map shows them in their context and that this helps him to take a more informed choice on which results are worth inspection and which ones not."*

*Implication 6.2: Tools should put search results into a meaningful context, so developers can take both quicker and better-informed decisions.*

The need for better presentation of search results has implications beyond the design of spatial visualizations. Work on presentation of search results goes beyond spatial maps [16], for example results can be presented as a graph. Poshyvanyk and Marcus [31] have taken one such approach (representing search results as a lattice) and applied it to source code search with promising results.

For our next prototype we plan to integrate search results into the package explorer view, just as is already done with compile errors (which are, from this point of view, just like the search results of a complex query that is run to find syntax errors). This planned feature addresses another implication of our study as well, as we have found that some developers establish a spatial memory of the package explorer view. It therefore makes sense to mark search results both on our map as well as in the explorer view.

*Observation 6.3: When interacting with the map, participants were attracted to isolated elements, rather than exploring clusters of closely related elements.*

We found that participants are more likely to inspect easily discernible elements on the map. They are more likely to notice and interact with an isolated island rather than with elements that are part of a larger continent. Unfortunately, it is exactly dense and strongly correlated clusters that contain the most interesting parts of the system! When investigating this issue, participants answered that *"those (isolated) elements looked more important as they visually stick out of the rest of the map."*

Also, when working with another system that had (unlike the present study) a large cluster in the middle surrounded by archipelagos on the periphery, we found that users started their exploration with isolated hills in the periphery, only then working their way towards the more dense cluster in the middle.

*Hypothesis 6.3/a: Developers avoided clusters of closely relates elements because they are difficult to identify and select on the map.*

All participants had difficulties to open files by clicking on the map. They had difficulties to select classes on the map when they are in a crowded cluster. They would click in the middle of a label, but often the labels are not centered, which is an unavoidable artifact of any labeling algorithm, and thus the clicks would open a different (unlabeled) class.

Codemap does provide tooltips, however participants did not use them. From observing their work it was obvious why: both students and professionals were working at such a speed that waiting for a tooltip to appear would have totally taken them out of their workflow.

*Observation 6.3/b: Participants rarely used Codemap to return to previously visited locations, instead using package explorer and "Open Type" to do so.*

Contrary to our assumptions, participants did not use the map to return to previously visited locations by recalling them from spatial memory. Some would use the map, but only for exposed classes that are easily recognizable and clickable. This observation is related to the previous one.

We found however some participants established a spatial memory of the package explore — and did so *in addition to their spatial model of the system's architecture!* For example, participant S would drill down with the explorer saying "let's open that class down there" or "there was this class up here." Over the course of the experiment he got quicker at navigating back to previously visited classes in the package explorer. Other participants, as for example participant T, relied on lexical clues and made extensive use of Eclipse's "Open Type" dialog to find their way back to previously visited classes.

Usability glitches will of course worsen the effect of (or might even be the main cause of) not using the map for navigation and revisiting classes. From this it follows that:

*Implication 6.3: The map's layout should be such that all elements are easily discernable and easy to click.*

Real estate on a computer screen is limited, and even more so in an IDE with all its views and panels. As tool builders we have limited space available for an embedded visualization. Given our goal of establishing a global layout we face the challenge of having to visualize all elements of a system in that very limited space.

The current implementation of Codemap has one level of scale only, which may yield crowded clusters where elements are placed just pixels apart. A zoomable map as provided by Code Canvas [11] addresses this issue.

The fact that we are attracted by elements that visually detach from other has two impacts: one is that we tend to look at isolated elements as being of low significance, the other being that it is hard to identify elements in the cluster. These impacts are very different, but can both be addressed in a common way. For instance, a threshold could be used to not show isolated elements at all, but only significant clusters. Alternatively, colors may be used to display isolated elements so that they do not draw our attention so readily.

*Observation 6.5: Participants used Codemap as if its layout were based on package structure — even though they were aware of the underlying topic-based layout.*

Developers assume that packages are a valid decomposition of the system and expect that the layout of the spatial visualization corresponds to the package structure. We found that clustering classes by topic rather than packages violates the "principle of least surprise." We observed that participants tended to interpret visual distance as a measure of structural dependencies — even though they were aware of the underlying lexical implementation!

Participants expected the layout to reflect at least some structural property. Most of them reacted surprised or confused when for example the classes of a package were not mostly in the same place. For example, Participant S reported in the wrap-up, *"this is a very useful tool but the layout does not make sense"*. Another participant stated during task 3 (*i.e.*, the architecture recovery) with confusion that *"the classes contained in packages are scattered on the map, it is not obvious what their spatial connection is."*

*Hypothesis 6.5: From the developers view, the predominant mental decomposition of a system is package structure.*

Given our reverse engineering background [26, 19, 13] we had come to distrust package decomposition, however it seems that developers like to rely on the packaging that other developers have made when designing the system.

One problem raised by research in re-packaging legacy systems is that packages play too many roles: as distribution units, as units of namespacing, as working sets, as topics, as unit of architectural components, etc. However, as an opposing point of view, we can relate packaging to the folksonomies of the Web 2.0, where users label elements with unstructured tags that are then exploited by other users to search for elements. In the same way, we could say that putting trust into a given package structure is a way of collaborative filtering. Developers assume that other developers had made the same choice as they would when packaging the system.

*Implication 6.5: The map layout should be based on code structure rather than latent topics only. However, non-structural data should be used to enrich the layout.*

When running the user study, it became quickly apparent that we should revise our initial assumption that lexical similarity is a valid dissimilarity metric for the spatial layout. This was the strongest feedback, and as is often the case in exploratory user studies, already obvious from watching the first professional participant for five minutes only. From all participants we got the feedback that they expect the layout to be structural and that our clustering by topics kept surprising them even after working with the map for almost two hours.

Still we think that spatial layouts that go beyond package structure are worthwhile. Therefore, we propose to enrich structure-based layout with non-structural data, such as design flaws. For future work, we are about to refine our layout algorithm based on that conclusion. The new layout is based on both lexical similarity and the ideal structural proximity proposed by the "Law of Demeter" (LOD). This is a design guideline that states that each method should only talk to its friends, which are defined as its class's fields, its local variables and its method arguments. Based on this we can defined an *idealized* call-based distance between software artifacts. Given a LOD-based layout, software artifacts are close to one another if they are supposed to call one another and far apart if they better should not call one another. Thus we get the desired property that visualizing call-graphs conveys meaningful arrow distances. On a LOD-based map, any long-distance-call has a diagnostic interpretation that helps developers to take actions: Long flow-map arrows indicate calls that possibly violate the "Law of Demeter".

## 7. THREATS TO VALIDITY

This section summarizes threats to validity. The study had a small sample size (3 students, 4 professionals) and might thus not be representative. We manually evaluated the data, results might thus be biased. Nevertheless, results are promising and running a pilot think-aloud study with a small user group is a state-of-the-art technique in usability engineering to learn learn about the reactions of users. Such pilot studies are typically used as feedback for further iteration of the tool and to assess the usefulness of its application [25].

## 8. RELATED WORK

In this section we discuss related tools and user studies. We selected tools that embed a spatial visualization of software in the IDE, and user studies that study how developers navigate in the IDE as well as studies of tools that group software artifacts by topic rather than structure.

Most closely related to the spatial approach studied in this paper is the work on spatial representation by DeLine [11, 7, 9, 10].

CodeCanvas by DeLine and Rowan [11] is a zoomable source editor that drops levels of details as you zoom out, up to the level of a UML diagram. CodeCanvas allows developers to rearrange the layout, to tear classes apart into smaller parts, and to filter by working sets and stack traces. CodeCanvas features two thematic maps, an interactive visualization of debug traces and spatial search results. A user study is not yet available.

Cherubini, Venolia *et al.* studied how developers use spatial representation of their code [7]. They asked a team to set up a wall-sized spatial map of their system, and observed them for three weeks. They found that developers use horizontal rows to lay out diagrams by layers, and vertical columns to group it by domain topic. These are the same two spatial decompositions we found in our study, *i.e.*, conceptual and structural. Also, they found that developers put utility classes aside, just as our study revealed.

CodeBubbles by Bragdon and Reiss *et al.* [5] breaks code into method fragments (called bubbles) and allows developers to arrange the code fragments spatially on a larger-than-screen map. Fragments do not overlap, and can be grouped into persistent working sets. The authors had been running a pilot study and reported that developers did not express concern over the absence of files, but rather felt that bubbles offer better editing. And further, that developers felt that spatial representation was very close to what is needed for visual explanation.

CodeBubbles differs from our approach in a fundamental aspect: developers create the spatial layout themselves as they add methods to the current working set. As a consequence there are as many spatial layouts of the system as there are working sets. By contrast our CODEMAP plug-in, as well as the CodeCanvas tool of DeLine [11], relies on one global layout for all programming tasks. In our present study we observed that some developers maintained more than one spatial model of the same system (one conceptual, one structural), we consider it thus promising future work to combine both approaches.

TeamTracks by Deline, Czerwinski and Robertson [10] assists developers using collaborative filtering of navigation history. They ran a user study where professional developers had to fix bugs. Developers that used TeamTracks showed better task completion. Moreover, they found that performance was better for locally focused features rather widely disperse features. This issue is addressed on our tool where we give visual evidence of quantity and dispersion of search results and call graphs, of which participants made good use in the feature location task.

CodeThumbnails by DeLine and Czerwinski *et al.* [9] embeds a global navigation map in the IDE. The map consists of a zoomed-out view of the source code text, similar to Eick's Seesoft [14]. In a user study they found that developers quickly established a spatial memory of the system. Results from our study confirm this observation.

Mylar (now known as Mylyn) by Kersten and Murphy [17] integrates a degree-of-interest (DOI) model in the IDE. Mylyn extends the IDE with a task view, and maintains a DOI model per task. Developers can this model to filter views by working set. This is similar to the clustering by topic on our spatial map, elements are grouped by conceptual rather than structural relation. They measured the approach in a user study and report an 15% productivity improvement, measured as "the number of keystrokes in the editor over the number of selections made in editor and views."

War Room Command Console by O'Reilly *et al.* [28] is a shared software visualization for team coordination. The system highlights individual team efforts by combining system structure and ongoing developer activity in order to report progress to management. They did a survey, and found that users felt the system was useful but found it hard to translate between viewpoints using the given spatial layout. Based on that, they propose a set of visualization techniques to overcome this limitation.

## 9. CONCLUSION

In this paper we presented an evaluation of spatial software visualization in the IDE. We embedded a prototype of the *Software Cartography* approach [22, 15, 20], the CODEMAP plug-in, in the Eclipse IDE and ran an exploratory user study which included both students and professionals.

Software maps are supposed to help developers with a visual representation of their software systems that addresses their spatial thinking and memory. The scenario of our user study was first contact with an unknown closed-source system. Results were as follows:

- Participants made good us of the map to inspect search results and call graph, they reported that the spatial visualization provided them with an immediate estimate of quantity and dispersion of search results.

- Participants found the layout of the map (which uses lexical information to cluster classes by topic) surprising and often confusing. This led to the revision of our initial assumption that lexical similarity is sufficient to lay out the cartographic map.

We made the following four main observations, and concluded from these the following implications:

- All participants used a form of spacial thinking to understand the system. It would be best to allow developers to rearrange the initial layout according to their spatial memory.

- Immediate estimate of quantity and dispersion of search results is useful and the map suits this well.

- Participants are distracted by isolated elements, which does not happen in textual/tabular representation. This is the drawback of visualization, which must find the right balance between the power of visualization and the pitfall of visualization. The map should be improved to mitigate that.

- The coexistence of two models for the software (one structural, one conceptual) causes some confusion. With the present map and implementation, participants were puzzled by non-structural nature of the map.

Developers intuitively expect that the map meets their mental model of the system's architecture. We observed that if this is not given, developers are not able to take advantage of the map's consistent layout. So for example, even though north/south and east/west directions had clear (semantic) interpretations in the map used for the user study, developers did not navigate along these axes.

However, even with the most perfect layout, developers might not be able to take advantage of the map if it elements are barely discernable, and thus difficult to inspect.

Based on the results of our user study, we conclude with the working hypothesis that *the map should incorporate structural information and be improved from point of usability* and that *we need more work to make two models (one structural, one conceptual) coexists without creating confusion.*

In order to achieve this, we propose the following changes to the Software Cartography approach:

- Compute the initial layout such that distance reflects structural correlation, since this is what the developers expect (*i.e.*, "principle of least astonishment").

- Use *anchored multi-dimensional scaling* for the layout such that developers may rearrange the map according to their spatial model of the system.

- To use architectural components as anchors for rearrangement of the map, since spatial thinking of developers is strongest at the architectural level.

- Improve the usability experience of inspecting and selecting of elements on the map, possibly using a zoomable user interface.

## Acknowledgements

We are grateful to Niko Schwarz and Erwann Wernli for their constructive comments that considerably contributed to this paper. We also express our thanks to all the fellow researchers, with whom Adrian Kuhn discussed parts of this work at the ICSE conference in Cape Town, 2010, for their input and inspiration. We gratefully acknowledge the financial support of the Swiss National Science Foundation for the project "Bringing Models Closer to Code" (SNF Project No. 200020-121594, Oct. 2008–Sept. 2010).